# Blockchain-Based Cloud Manufacturing: Decentralization

Ali Vatankhah BARENJI[a], Hanyang GUO[a], Zonggui TIAN[a], Zhi LI[a], W.M. WANG[a] and George Q. HUANG[b,1]

[a] *Guangdong University of Technology, Guangzhou, China*
[b] *University of Hong Kong, China*

**Abstract.** Recently, there has been growing interest in the field of cloud manufacturing (CM) amongst researchers in the manufacturing community. Cloud manufacturing is a customer-driven manufacturing model that was inspired by cloud computing, and its major objective was to provide ubiquitous on-demand access to services. However, the current CM architecture suffers from problems that are associated with a centralized based industrial network framework and third part operation. In a nutshell, centralized networking has had issues with flexibility, efficiency, availability, and security. Therefore, this paper aims to tackle these problems by introducing an ongoing project to a decentralized network architecture for cloud manufacturing which is based on the blockchain technology. In essence, this research paper introduces the blockchain technology as a decentralized peer to peer network for multiple cloud manufacturing providers.

**Introduction**

In the twentieth century, a collaboration between cloud computing and service-oriented structure as well as the internet of things(IoT) was identified as the key technological and developmental trends that were necessary for remolding the global manufacturing enterprises. Recently, cloud manufacturing (CM) has been fused into the manufacturing industry and has brought about a newer type of manufacturing infrastructure. CM is defined based on cloud computing, service-oriented architecture and the IoT, which has brought about on-demand service to the manufacturing industry [1]. Therefore, CM is defined as a new service-oriented manufacturing model that utilizes the internet and service platform to arrange the manufacturing resource. Also, it provides services according to the customers' demands [2]. Additionally, CM can be defined as a manufacturing paradigm that utilizes cloud computing and IoT to transfer the manufacturing resources and capabilities into the cloud environment as services which provide anything that is required by the customers and users. The main advantage of the CM is that it provides on-demand service for its customers. However, despite the potential benefits and advantages of the CM, it suffers from a centralized based network for communication purpose and third parts for managing [3]. As a result, these flaws within the centralized system tend to cripple the productivity of the CM;

---

[1] Corresponding Author, Mail: gqhuang@hku.hk.

these flaws include scalability issues and a broken communication model. Therefore, these flaws open the doors to more vulnerabilities which affect the system such as cyber-attacks and security problems. Also, most manufacturing industries have their own programs and business models and so, it would be quite cumbersome to develop a centrally organized network wherein all manufacturing industries would conform to the same set of predefined rules [4].

Just recently, a new peer to peer network was introduced as a blockchain and implemented in the economy as a type currency known as "bitcoin" [5]. Blockchain technology is a type of distributed, an electronic database that can hold any information and set rules on how this information will be updated. It develops as blocks and is linked via a hash. The hash is produced by running the contents of the block, in question, through a cryptographic hash function. An ideal cryptographic hash function can easily produce a hash for any input. However, it is difficult to use the hash to derive the input. Additionally, any changes in the original data would result in extensive and seemingly uncorrelated changes to the hash. This technology developed a distributed digital ledger of transactions that was shared amongst the nodes of a network as opposed to being stored on a central server [6]. The blockchain is capable of providing an effective solution to cloud manufacturing in terms of decentralized networks, such as effective collaboration, secure service exchange, scalability, and flexibility.

The purpose of this paper is to introduce an ongoing research on blockchain industrial network to cloud manufacturing; this will help fuse the advantages of decentralization into cloud manufacturing as well as cover up for the existing drawbacks associated to centralized cloud manufacturing. Therefore, we introduced our ongoing decentralized network architecture into cloud manufacturing which was based on the blockchain technology. In this paper, the proposed architecture contains two main parts, which will be explained in the next section; also, the properties and operation of the proposed decentralized cloud manufacturing are described in details with emphasis on how information is propagated through the blockchain network.

## 1. Related work

Blockchain technology is an immutable distributed ledger with a decentralized network that is cryptographically secured. The blockchain architecture is capable of sharing a ledger via the P2P repetition, which gets updated each time a block of the transaction(s) is agreed to be committed. Blockchain technology or distributed ledgers has burst onto the scene in the past few years as an important future technology. Though there is, as yet, neither a single nor internationally agreed upon definition of blockchain or distributed ledgers. They are often described as "an open source technology that supports trusted, immutable records of transactions stored in publicly accessible, decentralized, distributed, automated ledgers" [7].

We categorized the existing blockchain definition into three sections as follow:

- Business view: It is defined as a network for exchanging and transferring of value and money between willing and reciprocally approving members, by considering the warranting privacy and control of the data to participants.

- Legal view: Blockchain ledger transactions are authenticated, indisputable transactions, which do not require intermediaries or trusted third party legal entities.
- Technical view: It is defined as a distributed ledger of communications with ledger entries that references other information stores. In this view, members access only the parts of the ledger which are related to them. This security is created based on cryptography.

From the existing definition presented, we defined blockchain technology on cloud manufacturing as follow:

*Blockchain technology is a decentralized network for cloud manufacturing with a secured and distributed ledger that uses cryptography and smart contact for sharing services on the cloud between the providers and the end user. This network is secure, authenticated and verifiable based on the blockchain technology.*

As regards to the security and data sharing system, three types of blockchain systems exist; which can be defined as follow; *public blockchain* is public and anyone in the world can read and send the transactions, bitcoin, and other virtual currencies (which are developed based on this type). *Consortium blockchain* is partly private and is used consensus to control the nodes on the network. In this paper, we will use this type for developing decentralized cloud manufacturing. The last one is the *private blockchain* which is the similar to the centralized system for the developing an organization.

Recently, many researchers have introduced blockchain-based applications into several fields in order to try to take advantage of this technology within the various field. For example, Zhi Li et al [8]introduced the across-enterprises knowledge and services exchange framework based on blockchain technology. Sikorski et al[9] proposed a machine to machine communication via the blockchain technology and used a proof of concept to illustrate its implementation. Gallay Oliveir et al [10] introduced a novel platform for decentralized logistics based on the blockchain and the IoT. Blockchain and the IoT were used as basic building blocks, which when merged together, was capable of creating trusted and integrated platform for managing the logistics operations in a fully decentralized way.

The decentralized network is an entirely new concept in cloud computing. Decentralization in cloud computing plays a huge role in the implementation of the P2P computing services provided on the cloud network. Blockchain technology makes it possible to develop a decentralized cloud computing [11], in this respect, Gasper Skulj et al. [3] proposed an architecture for decentralized cloud manufacturing based on the autonomous work system. However, the proposed architecture's lack of standard modeling and standard communication brings a sense of lack of security and scalability to the system.

Based on the existing research on blockchain cloud computing, we developed blockchain cloud manufacturing that enables all evolving set of parties to maintain a safe, permanent, and tamper-proof ledger of connections without a central authority. In this system, service is not recorded centrally, instead, each cloud provider maintains a local copy of the ledger and service rendered. Each cloud provider uses a ledger to link up the blocks and send its service via a blockchain network to end user. Transactions are broadcasted and recorded by each member in the blockchain based cloud manufacturing network. When a new cloud provider wants to update the system, a block is proposed. The participants in the network then agree upon a single valid copy of this block according to a consensus mechanism. Once a block is collectively accepted, it is practically impossible to change it or remove it.

## 2. Blockchain Cloud Manufacturing Architecture

The basic components of blockchain cloud manufacturing are illustrated in Fig 1, which follows the existing cloud manufacturing components. All the components of blockchain cloud manufacturing architecture together with their functionalities are defined in such a way that it allows for simplicity, speed, and efficiency (i.e. when a new cloud provider enters in the architecture). Another key characteristic is the communication between the end users and CM which is realized through a blockchain network(P2P). As a result, there are many advantages that are observed such as Scalability, blockchain is a P2P network which permits the integration of many interconnected machines. Security, blockchain provided ledged and smart contact for communication purpose. Fault tolerance, since there is no centralized system, some nodes failing will not result in a total system shutdown (i.e. the others can work independently). Flexibility, each cloud manufacturing can operate independently or in a dependent manner. The proposed architecture consists of two main parts. Firstly, the user part, which is the end user layer on the CM. However, this isn't the main objective of this paper and so, will be discussed briefly. The main part of this architecture is the blockchain cloud manufacturing which consists of three main layers, namely, the core layer, P2P network layer (blockchain network) and cloud manufacturing layer.

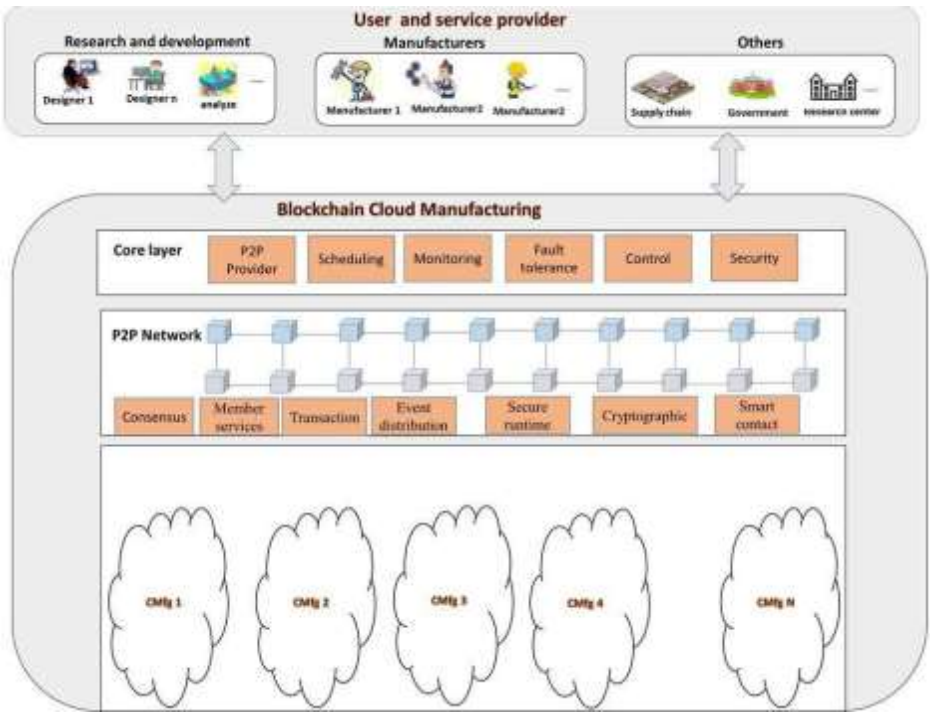

**Figure 1.** The architecture of Blockchain Cloud Manufacturing.

*2.1. The end user*

This part provides the service of interaction for the end users, which can be implemented by the use of the graphical interfaces (GUI) or workflow management system [12].The end users can use services from the existing manufacturing systems on the architecture via the blockchain technology as P2P network and can be chosen among the available services. This part used a demand control service which is provided by the blockchain network for the end users. This controller is responsible for collecting end user request and sending it to the core layer as input data via the blockchain network. Also, this part is responsible for providing the user interface that runs the applications.

*2.2. Blockchain cloud manufacturing*

This part has three main layers, namely, the core layer, P2P network or blockchain network and the cloud manufacturing systems. These layers are explained as follows.

*Core layer*

This layer plays the role of a manager within the system by managing all the other layers on the system. Its main responsibility is to provide a blockchain network and support the P2P network. However, as a sub-responsibility, it helps in identification of new services or cloud providers, scheduling of the system, the control, and maintenance of the system. Core layer provides the services for blockchain networks such as the block generation service, the motivation of network service, management service and etc. At the same time, it is responsible for providing services to others parts such as; demand control service, discovery service, SLA service, monitoring service, security service and etc.

*Blockchain network or P2P network*

Blockchain network plays the main role in the system and acts as a distributed network in the architecture. Therefore, Blockchain network is developed based on the number of nodes, each of which has a local copy of the ledger. In the proposed system, the nodes belong to different cloud manufacturing providers or users. Nodes are responsible for creating a secure communication between each other to gain the service or contract of the leger and don't require a central power to coordinate and validate any transactions. The process of gaining a service is called consensus, and there are a number of different algorithms. End users send transaction requests to the blockchain cloud manufacturing in order to use the specific service in the chain; which is designed to provide a service to the end user with other centralized power. Once a transaction is completed, a record of the transaction is added to the ledgers and can never be removed. This property of the proposed system follows the immutability property of the blockchain. In the blockchain, cloud manufacturing cryptography guarantees the security of the service exchange between the users and the service providers. As a matter of fact, it ensures that the ledger cannot be altered, except by the addition of new transactions. Cryptography provides integrity to the messages from users or between nodes and ensures that the operations are only performed by the authorized entities concerned. Therefore, the proposed blockchain network which acts as a P2P, contains five major parts, namely the chained data block, network part, agreement part, motivation part and the contact part. Table 1 summarizes the purpose and components of each part.

**Table 1.** Components of Blockchain Network.

| Part | Responsibility and Purpose | Main modules |
|---|---|---|
| Chained data block | It is responsible for providing the chained data blocks to the system with the related techniques, which is used for transferring the data to the HD. | Hash Algorithms, Asymmetric Encryption and Merkle Tree and Time Stamping. |
| Network part | It specifies the mechanisms of the distributed networking and represents the data forwarding peer to peer communication as well as the verification of the communication. It is responsible for verifying the legality of the broadcasted message as well as the management of the peer connection. | Data Forwarding, Peer Management, and Data Verification. |
| Agreement part | Also called the consensus or handshake part. It provides decentralized communication within the network and helps to establish trust between the unknown users in the communication environment. | PoW based algorithm, Proof of Stake, Delegated Proof of Stake algorithm and Proof of Movement algorithm. |
| Motivation part | It is responsible instilling a sense of motivation on the network by generating a reward for new blocks i.e. it appreciates the effort made towards data verification. | Crypto Data. |
| Contact part | It is responsible for serving as activators for the static data by using various scripts, algorithms and smart contracts. | Script and algorithms. |

*Cloud manufacturing provider layer*

This layer consists of many cloud manufacturing providers, the core part ensures a unified view of the cloud, which allows users to use all the available service via a P2P network. Each cloud manufacturing can be connected to this platform as a service provider for the customer. Each cloud manufacturing follows the standard cloud architecture that can be considered as either a public or private cloud. The main advantage of this architecture is decentralization, meaning that each cloud provider is responsible for its own cloud environment and can collaborate and cooperate with other providers via the blockchain network.

## 3. Typical characteristic of the blockchain cloud manufacturing

*3.1. On-demand secure services*

Blockchain cloud manufacturing was proposed based on the existing on-demand IT service (storage, network and etc.) such as the CM or smart manufacturing. However, the main advantages of the proposed system was the use of the P2P for data sharing and service exchange i.e. throughout the manufacturing life cycle via the blockchain technology; this helped in providing on-demand secure service for the whole system by taking into account that the BCN's originality of data was defined in a deterministic way. This is done so as to know which user sent the data to the system and where the user is. Therefore, the proposed system provides a secured platform for on-demand secure service.

*3.2. Supports secure data exchange*

Large data is the foundation on which CM was built. Due to the usage of the IoT and computer-based machines in the manufacturing system, there has been a large number of data generated in the new manufacturing paradigms; these data can be sensed and collected over the internet and cloud to be processed. Data exchange plays a crucial role of service exchange in the new manufacturing paradigms; in this respect, secured data has become one of the important issues that need to be taken care of. As mentioned previously, data sharing in the existing system suffers from security and scalability issues due to a centralized system. In the proposed architecture, each cloud manufacturing provider has one data security system that is not controlled by the centralized node. Moreover, the blockchain network played an important role in improving the security of the data sharing between the end users and the providers, which is based on the smart contact and cryptocurrency.

*3.3. Effective collaboration*

In reality, manufacturing industries have their own programs and business models and so, it is quite cumbersome to develop a central cloud manufacturing structure. The proposed architecture provides a remedy for this as it is also a decentralized collaboration platform (i.e. it collaborates with the different cloud manufacturing providers) which is guaranteed by the blockchain network communication. Therefore, in this architecture, manufacturing industry can effectively collaborate with other industry without changing their business models.

**4. Execution of job in the proposed architecture**

In the proposed architecture, when the end user wants to use the service, the first step is to launch the web interface and then, develop the wallet after the specific end user requests a service with informed details of the application to be executed. After the information has been entered by the end user layer with collaboration with the core layer created by the block, the information is encapsulated in the block for the service request from the cloud manufacturing providers that are available. Figure 2 shows how the information is encapsulated into the block and broadcasted on the network. After sending this information to the blockchain network core layer, the management and scheduling of all system are done and then, the related request is sent to the cloud manufacturing providers. The cloud manufacturing providers can send service via a P2P network to the customer and the end user can utilize this service based on the blockchain network. In the case of collaboration, any cloud manufacturing offer request can either be public or private to the system. Moreover, other existing cloud manufacturing can collaborate via a distributed secure data sharing protocol in the proposed architecture.

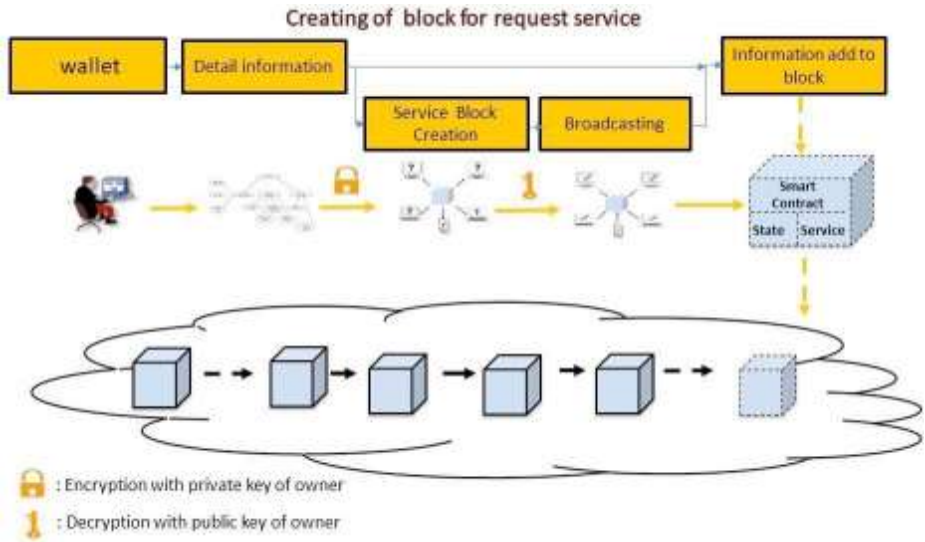

**Figure 2.** Encapsulation of data on the developed block

## 5. A case study

For a better understanding of proposed architecture, we are going to the explained illustrative example, in this case, we consider a customer who wants to request a matching as a service, As simple scenario customer want to produce a simple part, the part needs to have ellipse in the middle of cube which can be done with CNC milling machine. Therefore customer after singing into a web application and creating a customer block for broadcasting on the BCN, the information related to service encapsulated on the block via the end user such as a characteristic of the product, due date, and price this information was guaranteed by the encryption process. The service providers read the offers, analyses them based on provided information and send an offer to the customer (product cost and due date) in this process smart contact between the end user and service providers are also created. The customer reads the offers than decided to select service (minimum cost and due date). When customer accepted an offer from the service provider, it is executed as an atomic exchange. And developed smart contact between the user and provider guarantee the service quality and payment of the service.

## 6. Conclusions

Cloud manufacturing is quickly emerging as a new manufacturing paradigm as well as an integrated technology that is capable of revolutionizing the manufacturing industry towards a service-oriented, highly collaborative and innovative manufacturing. This paper first discussed the existing drawbacks of cloud manufacturing and highlighted the centralized based cloud manufacturing problems in details. For a viable solution to this problem and to improve the existing cloud manufacturing, a decentralized cloud manufacturing based on the peer to peer network that was developed via the blockchain technology, was proposed. We introduced the decentralized cloud manufacturing architecture and highlighted the advantages of this architecture. In order to develop and implement the proposed system in the future, we will first focus on the encapsulation of the service into the block and afterward, we will focus on the proposed model system's capability as well as the virtualization of the cloud manufacturing in the blockchain network.

**References**

[1] X. Xu, From cloud computing to cloud manufacturing, *Robotics and computer-integrated manufacturing*, 2012, 28, pp. 75–86.
[2] F. Tao et al., Cloud manufacturing: a computing and service-oriented manufacturing model. *Proceedings of the Institution of Mechanical Engineers, Part B: Journal of Engineering Manufacture*, 2011, 225(10), pp. 1969--1976..
[3] G. Škulj, et al., Decentralised network architecture for cloud manufacturing, *International Journal of Computer Integrated Manufacturing*, 2017. 30(4-5): p. 395-408.
[4] A.V. Barenji, et al., A dynamic multi-agent-based scheduling approach for SMEs, *The International Journal of Advanced Manufacturing Technology*, 2017, 89(9-12), pp. 3123-3137.
[5] S. Underwood, Blockchain beyond bitcoin. Communications of the ACM, 2016. 59(11): p. 15-17.
[6] M. Swan, Blockchain: Blueprint for a new economy, O'Reilly Media, Inc., 2015.
[7] C. Cachin, Architecture of the Hyperledger blockchain fabric. In *Workshop on Distributed Cryptocurrencies and Consensus Ledgers*. 2016.
[8] Li Z, Barenji AV, Huang GQ. Toward a blockchain cloud manufacturing system as a peer to peer distributed network platform. Robotics and Computer-Integrated Manufacturing. 2018 Dec 1;54:133-44.
[9] J.J. Sikorski, J. Haughton, and M. Kraft, Blockchain technology in the chemical industry: Machine-to-machine electricity market, *Applied Energy*, 2017, 195, pp. 234-246.
[10] Gallay, O., et al. A peer--to--peer platform for decentralized logistics, in Proceedings of the Hamburg International Conference of Logistics (HICL). 2017. epubli.
[11] Liang, X., et al. Provchain: A blockchain-based data provenance architecture in a cloud environment with enhanced privacy and availability. In *Proceedings of the 17th IEEE/ACM International Symposium on Cluster, Cloud and Grid Computing*, 2017, IEEE Press.
[12] H. Saldanha et al., Towards a hybrid federated cloud platform to efficiently execute bioinformatics workflows, in *BioInformatics*, 2012, InTech.
[13] Vatankhah Barenji A, Hashemipour M. Real-Time Building Information Modeling (BIM) Synchronization Using Radio Frequency Identification Technology and Cloud Computing System. *Journal of Industrial and Systems Engineering*. 2017 Apr 19;10:61-8.
[14] Barenji AV. Cloud-based manufacturing execution system: case study FMS. International Journal of Industrial and Systems Engineering. 2018;30(4):449-67.